\documentclass[pra,showpacs,nofootinbib]{revtex4}
\usepackage{hyperref}
\usepackage{amsmath}
\usepackage{amssymb}
\usepackage{amsthm}
\usepackage{graphics}
\usepackage{graphicx}
\usepackage{enumerate}
\usepackage{color}
\usepackage[usenames,dvipsnames,svgnames,table]{xcolor}

\long\def\ca#1\cb{}

\def\bra#1{\langle#1|}

\def\inpd#1#2{\langle#1|#2\rangle }
\def\ket#1{|#1\rangle }

\def\Tr#1{\textrm{Tr}\left(#1\right)}

\def\HC{{\cal H}}
\def\IC{{\cal I}}

\def\KC{{\cal K}}
\def\LC{{\cal L}}
\def\MC{{\cal M}}

\def\SC{{\cal S}}

\def\endproof{{\hspace{\stretch{1}}$\blacksquare$}}

\newtheorem{thm1}{Theorem}
\newtheorem{thm2}[thm1]{Theorem}

\newtheorem{lem1}[thm1]{Lemma}
\newtheorem{cor1}[thm1]{Corollary}
\newtheorem{lem2}[thm1]{Lemma}

\begin{document}
\title{A class of unambiguous state discrimination problems achievable by separable measurements but impossible by local operations and classical communication}
\author{Scott M. Cohen}
\email{cohensm52@gmail.com}
\affiliation{Department of Physics, Portland State University, Portland OR 97201}

\begin{abstract}
We consider an infinite class of unambiguous quantum state discrimination problems on multipartite systems, described by Hilbert space $\HC$, of any number of parties. Restricting consideration to measurements that act only on $\HC$, we find the optimal global measurement for each element of this class, achieving the maximum possible success probability of $1/2$ in all cases. This measurement turns out to be both separable and unique, and by our recently discovered necessary condition for local quantum operations and classical communication (LOCC), it is easily shown to be impossible by any finite-round LOCC protocol. We also show that, quite generally, if the input state is restricted to lie in $\HC$, then any LOCC measurement on an enlarged Hilbert space is effectively identical to an LOCC measurement on $\HC$. Therefore, our necessary condition for LOCC demonstrates directly that a higher success probability is attainable for each of these problems using general separable measurements as compared to that which is possible with any finite-round LOCC protocol.
\end{abstract}

\date{\today}

\maketitle
\section{Introduction}\label{sec1}
In a recent paper \cite{myExtViolate1}, we proved a necessary condition (reproduced as Theorem~\ref{thm1}, below) that a multi-party quantum measurement can be implemented by local operations and classical communication (LOCC) in any finite number of rounds of communication. It is easily seen that such measurements must be separable---that is, the measurement operators must all be tensor products---and our Theorem~\ref{thm1} provides a strong, and quite general, constraint on the set of product operators representing any measurement implemented by finite-round LOCC. We also showed that the condition of Theorem~\ref{thm1} is extensively violated by separable measurements, a violation limited only by the size of the system, as measured by the number of parties involved.

Despite the generality of Theorem~\ref{thm1}, we were unable, at the time of writing, to provide examples of separable measurements having obvious practical interest, and which violate the conditions of that theorem. A reasonable criticism, then, was that the theorem was ``primarily of mathematical value with physical implications wanting" \cite{thanksRef}. Here, we remedy this deficiency by providing an infinite class of physically motivated examples where the theorem can be directly used to demonstrate the LOCC-impossibility of these specific operational tasks, each of which can, nonetheless, be implemented by separable measurements.

Our examples involve the optimal unambiguous discrimination of quantum states, a subject pioneered by Ivanovic \cite{Ivanovic}, Dieks \cite{Dieks}, and Peres \cite{Peres}. This is one method of extracting information from non-orthogonal states, wherein due to this non-orthogonality, the information cannot be obtained perfectly. There are numerous scenarios that involve the extraction of information under such conditions, including quantum cryptography and quantum key distribution \cite{QCryptUSD}. It is therefore a subject of considerable significance in quantum information processing, with implications for both theory and experiment, and its study remains robust to this day \cite{RoaKlimov,BergouPRL,ChitDuanHsieh,Waldherr}. 

In the scenario of unambiguous state discrimination, a quantum system is prepared in one of a given set of states, and the aim is to perform measurements on that system in order to determine in which state it was prepared. It is required that the error probability is zero---one can never guess one state when it happens to be another---which means that when the states are not mutually orthogonal, there must be an inconclusive outcome, one for which the given state remains unknown.

Chefles \cite{Chefles} has shown that the states in the given set can be unambiguously discriminated if and only if they are linearly independent, and then the measurement involves the reciprocal set of states, see below. When the states form a symmetric set and the a priori probabilities are all equal, then an optimal measurement---one achieving the maximum possible success probability---was obtained in \cite{CheflesBarnett}. Later, Eldar \cite{Eldar} showed that the problem of finding an optimal measurement for an arbitrary set of linearly independent pure states can be formulated as a semidefinite programming problem.

We will assume that the quantum system under consideration is made up of $P$ spatially separated parts, and that the separate parties utilize LOCC in order to discriminate the states. Chefles \cite{CheflesLOCC} found a condition, valid for both separable measurements and for LOCC, which is necessary and sufficient that a set of states can be unambiguously discriminated. The equivalence of LOCC and the full set of separable measurements for this question is not obvious, even though every LOCC is also separable \cite{Rains}. The reason is the existence of separable measurements that cannot be implemented by LOCC, a discovery first made in \cite{Bennett9}. Of course, this result of \cite{CheflesLOCC} does not say that for unambiguous state discrimination, use of the full set of separable measurements is equivalent to using only LOCC, because there is still the question of finding an optimal measurement. Along these lines, it was shown in \cite{Koashi} that the success probability with general separable measurements can exceed that for LOCC for a pair of two-qubit states, one pure and the other mixed. As far as we are aware, this is the only known example of a separation between separable measurements and LOCC for unambiguous state discrimination. Here, we provide an infinite set of new examples showing such a separation, all of which only involve pure states, including one case involving two qubits. For these examples, we use Theorem~\ref{thm1} to show that LOCC cannot achieve as high a success probability as is possible with separable measurements. This then accomplishes a main goal of this paper, which is to demonstrate the utility of Theorem~\ref{thm1}.

Consider a multipartite system of $P$ parts, described by Hilbert space $\HC=\HC_1\otimes\cdots\otimes\HC_P$ of overall dimension $D$, and a separable measurement on $\HC$ consisting of $N$ measurement operators $\KC_j=\KC_j^{(1)}\otimes\cdots\otimes\KC_j^{(P)}$ satisfying $\KC_j\ge0$ and $\sum_j\KC_j=I$, with $I$ the identity on $\HC$. Following the ideas in \cite{myExtViolate1}, we consider the convex cones generated by the set of local operators $\{\KC_j^{(\alpha)}\}$, for each $\alpha$. As the number of operators is finite, these are polyhedral cones, having a finite number of extreme rays.\footnote{A ray is a half-line of the form $\{\lambda\hat\KC_j^{(\alpha)}\vert\lambda\ge0\}$. An extreme ray of a convex cone is a ray that lies in the cone but cannot be written as a positive linear combination of other rays in that cone.} Let us count the distinct extreme rays in the convex cone generated by the set of local operators $\{\KC_j^{(\alpha)}\}$, for each party $\alpha$, and define this number to be $e_\alpha$. Then, the following theorem was proved in \cite{myExtViolate1}.
\begin{thm1} \label{thm1}For any finite-round LOCC protocol of $P$ parties implementing a separable measurement corresponding to the $N$ distinct positive product operators $\{\KC_j=\KC_j^{(1)}\otimes\ldots\otimes\KC_j^{(P)}\}_{j=1}^N$, it must be that
\begin{align}\label{eqn00}
\sum_{\alpha=1}^P e_\alpha\le2(N-1),
\end{align}
where $e_\alpha$ is the number of distinct extreme rays in the convex cone generated by operators $\{\KC_j^{(\alpha)}\}_{j=1}^N$, and the sum includes only those parties for which at least one of these local operators is not proportional to the identity.
\end{thm1}
In \cite{myExtViolate1}, we presented separable measurements consisting of a set of product operators $\{\Psi_k\}_{k=1}^N$ for every $D$ and prime $N>D$, for which the upper bound in this theorem is violated maximally, satisfying $\sum e_\alpha=PN$, thus demonstrating a very strong difference between separable measurements and LOCC. In Section~\ref{sec2}, we use these same operators to construct sets of states for which the optimal global measurement for unambiguous state discrimination is separable and (for present purposes, effectively) unique, see Theorem~\ref{thm2}, and which cannot be implemented by finite-round LOCC, a result that follows immediately from Theorem~\ref{thm1}. Theorem~\ref{thm2} states further that the optimal probability of success, which is achievable by a separable measurement, cannot be achieved using finite-round LOCC, even when applied to an enlarged Hilbert space. In Section~\ref{sec3}, we give a proof of Theorem~\ref{thm2}, and then we offer our conclusions in Section~\ref{sec4}.

\section{Separable measurements that are strictly better than LOCC}\label{sec2}
Consider any prime number $N\ge5$ and a multipartite system having overall dimension $D=N-1$. The number of parties $P$ can be chosen in any way consistent with the prime factorization of $D$---this choice is generally not unique, but it is unimportant for our present purposes. Let $\HC_\alpha$ be the Hilbert space describing party $\alpha$'s subsystem, and the overall Hilbert space is then $\HC=\HC_1\otimes\HC_2\otimes\ldots\HC_P$. Define states
\begin{align}\label{eqn101}
\ket{\Psi_j}=\ket{\psi_j^{(1)}}\otimes\ldots\otimes\ket{\psi_j^{(P)}},~j=1,\ldots,N,
\end{align}
\noindent with
\begin{align}\label{eqn1}
\ket{\psi_j^{(\alpha)}}=\frac{1}{\sqrt{d_\alpha}}\sum_{m_\alpha=0}^{d_\alpha-1}e^{2\pi \textrm{i}jp_\alpha m_\alpha/N}\ket{m_\alpha},
\end{align}
\noindent where $d_\alpha$ is the dimension of $\HC_\alpha$, with parties ordered such that $d_1\le d_2\le\cdots\le d_P$, and overall dimension $D=d_1d_2\cdots d_P$. 
Here, $p_1=1$ and for $\alpha\ge2$, $p_\alpha=d_1d_2\cdots d_{\alpha-1}$, and $\ket{m_\alpha}$ is the standard basis for party $\alpha$. It was shown in \cite{myExtViolate1} that
\begin{align}\label{eqn2}
I=\frac{D}{N}\sum_{j=1}^N\Psi_j,
\end{align}
where $\Psi_j=\ket{\Psi_j}\bra{\Psi_j}$.

We will choose $D$ of the $\ket{\Psi_j}$ and then show they are a basis of the full space. First note that diagonal unitary
\begin{align}
U^{(\alpha)}=\sum_{m_\alpha=0}^{d_\alpha-1}e^{2\pi \textrm{i}p_\alpha m_\alpha /N}\ket{m_\alpha}\bra{m_\alpha}\notag
\end{align}
permutes the states $\ket{\psi_j^{(\alpha)}}$. That is, $U^{(\alpha)}\ket{\psi_j^{(\alpha)}}=\ket{\psi_{j+1}^{(\alpha)}}$, and we have set $\ket{\psi_{N+1}^{(\alpha)}}=\ket{\psi_1^{(\alpha)}}$. Therefore, $U=U^{(1)}\otimes\cdots\otimes U^{(P)}$ permutes the $\ket{\Psi_j}$, showing that the latter $N$ states are a symmetric set. (Note, however, that the chosen $D=N-1$ states are \textit{not} a symmetric set.) As a consequence, it does not matter which of the $N$ states we omit in choosing a basis of the full $D$-dimensional space---any conclusions reached by omitting one could equally well have been reached by omitting any other---so without loss of generality, we will choose to omit $\ket{\Psi_1}$.

Let us then define two sets of states,
\begin{align}\label{eqn102}
\SC_\Psi=\{\ket{\Psi_j}\}_{j=2}^N,
\end{align}
and
\begin{align}\label{eqn103}
\SC_\Phi=\{\ket{\Phi_j}\}_{j=2}^N,\end{align}
\noindent each set reciprocal to the other. [Given \eqref{eqn101} and \eqref{eqn1}, this reciprocity is how the $\ket{\Phi_j}$ are to be determined, see \eqref{eqn4} below.] Our aim is to unambiguously discriminate $\SC_\Phi$. Existence of a reciprocal set of states requires that the original set is linearly independent, so we must demonstrate that the $D$ states of $\SC_\Psi$ possess this property. Actually, the following lemma is more general than what we need.
\begin{lem1}\label{lem1}
Given a prime number $N$ and any $D\le N$, any subset $\IC\subseteq[1,\ldots,N]$ of $D$ or fewer of the states $\ket{\Psi_j}$, defined in \eqref{eqn101}, constitutes a linearly independent set.
\end{lem1}
\noindent\proof Consider
\begin{align}\label{eqn3}
0&=\sum_{j\in\IC}c_j\ket{\Psi_j}\notag\\
&=\sum_{m_1=0}^{d_1-1}\cdots\sum_{m_P=0}^{d_P-1}\left(\sum_{j\in\IC}c_je^{2\pi \textrm{i}j\sum_\alpha p_\alpha m_\alpha/N}\right)\ket{m_1\ldots m_P}\notag\\
\Longleftrightarrow0&=\sum_{j\in\IC}c_je^{2\pi \textrm{i}j\sum_\alpha p_\alpha m_\alpha/N}~\forall{m_1,\ldots,m_P}.
\end{align}
By an argument similar to that following Eq.~(6) of \cite{myExtViolate1}, one sees that $\sum_\alpha p_\alpha m_\alpha$ takes on each value ranging from $0$ to $D-1$, and each of these values corresponds to a unique set of the indices, $m_\alpha$. Choosing any $\vert \IC\vert$ values of $k=k\left(\{m_\alpha\}\right)=\sum_\alpha p_\alpha m_\alpha$ from \eqref{eqn3}, we can represent these $\vert \IC\vert$ constraints as $M\vec{c}=0$, where the $j$th component of $\vec{c}$ is $c_j$, and $\vert \IC\vert\times\vert \IC\vert$ matrix $M$ has components $M_{kj}=e^{2\pi \textrm{i}jk/N}$. It is clear that $M$ is a sub-matrix of the $N\times N$ matrix $\Omega=(\omega^{jk})_{j,k=0}^{N-1}$, with $\omega=e^{2\pi \textrm{i}/N}$ a primitive root of unity. Then, by Chebotar\"{e}v's theorem on roots of unity \cite{ChebotarevProof}, $M$ is invertible. This implies that $c_j=0~\forall{j\in\IC}$ and that the set $\{\ket{\Psi_j}\}_{j\in\IC}$ is linearly independent, which completes the proof.\hspace{\stretch{1}}$\blacksquare$

Given states $\ket{\Psi_j}$, the reciprocal states $\ket{\Phi_j}$ are defined by the relations 
\begin{align}\label{eqn4}
\inpd{\Psi_k}{\Phi_j}=\delta_{jk}\inpd{\Psi_j}{\Phi_j}~\forall{j,k=2,\ldots,N}.
\end{align}
The set of states $\SC_\Phi$, given with a priori probabilities $\eta_j$, can be unambiguously discriminated \cite{Chefles} by the measurement $\MC\left(\{w_j\}\right)$ consisting of operators $\{w_j\Psi_j\}_{j=2}^N$ and one additional ``failure" operator,
\begin{align}\label{eqn5}
\Pi_f&=I-\sum_{j=2}^Nw_j\Psi_j.
\end{align}
To find an optimal measurement, the weights $w_j$ are chosen so as to minimize the probability of failure,
\begin{align}\label{eqn6}
Pr(f)=\sum_{j=2}^N\eta_j\Tr{\Pi_f\Phi_j},
\end{align}
with $\Phi_j=\ket{\Phi_j}\bra{\Phi_j}$.

Note that since the $D$ chosen $\ket{\Psi_j}$ are linearly independent and a basis of the full Hilbert space $\HC$, their reciprocal states $\ket{\Phi_j}$ are also linearly independent and a basis of $\HC$. Therefore, since $\inpd{\Psi_k}{\Phi_j}=0~\forall{j\ne k}$, if for some $k$ we replace $\ket{\Phi_k}$ by $\ket{\Psi_k}$ in the basis formed by the states in $\SC_\Phi$, we will still have a basis of $\HC$. This means that if we exclude $\ket{\Phi_k}$ from the set $\SC_\Phi$, the only positive operator that annihilates all of the remaining states is $\Psi_k$, up to multiplicative factors. Hence, we have the following corollary to Lemma~\ref{lem1}.
\begin{cor1}\label{cor1}
The only positive operators acting on $\HC$ that will unambiguously identify $\ket{\Phi_k}$ are those proportional to $\Psi_k$ defined in \eqref{eqn101}. This implies that the only such measurements unambiguously discriminating the set $\SC_\Phi$ are those of the form $\MC\left(\{w_j\}\right)$ defined above \eqref{eqn5}, and the only freedom available for optimizing these measurements lies in the choice of the $w_j$.
\end{cor1}

In the next section, we will prove the following theorem.
\begin{thm2}\label{thm2}
If a priori probabilities $\eta_j=1/D~\forall{j}$, then the optimal global measurement, $\MC_{opt}$, for unambiguous discrimination of the set of states that is reciprocal to any $D=N-1$ of the states defined in \eqref{eqn101} is 
\begin{enumerate}[(i)]
  \item separable;
  \item unique, when restricting to measurements that act only on $\HC$;
  \item consists of measurement operators $D\Psi_j/N, j=1,\ldots,N$;~and
  \item achieves $Pr(f)=0.5$;
\end{enumerate}
In addition, $Pr(f)>0.5$ for this task when using any finite-round LOCC protocol.
\end{thm2}
\noindent The last statement in this theorem requires the following lemma, which is proved in the appendix.
\begin{lem2}\label{lem2}
 Given a multipartite system of $P$ parties described by Hilbert space $\HC$, consider any enlargement of $\HC$ to $\HC^\prime$. Then, for any LOCC protocol $\LC$ implementing a measurement on $\HC^\prime$ that involves input states that are supported only on $\HC$, there exists an effectively identical LOCC measurement on $\HC$ which accomplishes precisely what is accomplished by $\LC$.
\end{lem2}

Before moving on to the proof of Theorem~\ref{thm2}, let us give an explicit example, providing expressions for the states, $\ket{\Phi_j}$, in the case of two qubits with $N=5$. These four states are
\begin{align}\label{eqn12}
\ket{\Phi_2}&=\frac{1}{\sqrt{5+\sqrt{5}}}\left(-e^{-2\pi \textrm{i}/5}\ket{00}+(1+e^{-2\pi \textrm{i}/5})\ket{01}-(1+e^{2\pi \textrm{i}/5})\ket{10}+e^{2\pi \textrm{i}/5}\ket{11}\right),\notag\\
\ket{\Phi_3}&=\frac{1}{\sqrt{5+\sqrt{5}}}\left(\frac{e^{2\pi \textrm{i}/5}}{2\cos(2\pi/5)}\ket{00}-\ket{01}-e^{\pi i/5}\ket{10}+(1+e^{-2\pi \textrm{i}/5})\ket{11}\right),\notag\\
\ket{\Phi_4}&=\frac{1}{\sqrt{5+\sqrt{5}}}\left((1+e^{2\pi \textrm{i}/5})\ket{00}-e^{\pi i/5}\ket{01}+e^{\pi i/5}\ket{10}-(1+e^{2\pi \textrm{i}/5})\ket{11}\right),\notag\\
\ket{\Phi_5}&=\frac{1}{\sqrt{5+\sqrt{5}}}\left(\ket{00}+\frac{e^{2\pi \textrm{i}/5}}{2\cos(2\pi/5)}\ket{01}+(1+e^{2\pi \textrm{i}/5})\ket{10}-e^{-2\pi \textrm{i}/5}\ket{11}\right).
\end{align}
We note that these states are all entangled, their reduced density matrices having Von Neumann entropy approximately equal to $0.3$, the same for all four states. The optimal measurement to unambiguously discriminate this set of states is given by operators $D\Psi_j/N,~j=1,2,3,4,5$, with
\begin{align}\label{eqn13}
\ket{\Psi_1}&=\frac{1}{2}\left(\ket{00}+e^{4\pi i/5}\ket{01}+e^{2\pi \textrm{i}/5}\ket{10}+e^{-4\pi i/5}\ket{11}\right),\notag\\
\ket{\Psi_2}&=\frac{1}{2}\left(\ket{00}+e^{-2\pi \textrm{i}/5}\ket{01}+e^{4\pi i/5}\ket{10}+e^{2\pi \textrm{i}/5}\ket{11}\right),\notag\\
\ket{\Psi_3}&=\frac{1}{2}\left(\ket{00}+e^{2\pi \textrm{i}/5}\ket{01}+e^{-4\pi i/5}\ket{10}+e^{-2\pi \textrm{i}/5}\ket{11}\right),\notag\\
\ket{\Psi_4}&=\frac{1}{2}\left(\ket{00}+e^{-4\pi i/5}\ket{01}+e^{-2\pi \textrm{i}/5}\ket{10}+e^{4\pi i/5}\ket{11}\right),\notag\\
\ket{\Psi_5}&=\frac{1}{2}\left(\ket{00}+\ket{01}+\ket{10}+\ket{11}\right).
\end{align}
The failure operator in this case is $\Pi_f=D\Psi_1/N$. We now proceed to the proof of Theorem~\ref{thm2}.

\section{Proof of Theorem~\ref{thm2}}\label{sec3}
We begin by noting that Lemma~\ref{lem2} applies to any number of parties, including when there is only one. Therefore, if we find an optimal global measurement under the restriction that it acts only on $\HC$, then this measurement is also optimal without such a restriction. Then, according to Corollary~\ref{cor1}, we can find an optimal measurement by considering $\MC\left(\{w_j\}\right)$, defined by \eqref{eqn5} and the sentence which precedes it, and then minimizing the probability of failure over all choices of the weights, $w_j$. Inserting \eqref{eqn2} into \eqref{eqn5}, we have
\begin{align}\label{eqn16}
\Pi_f&=\frac{D}{N}\sum_{j=1}^N\Psi_j-\sum_{j=2}^Nw_j\Psi_j,\notag\\
	&=\frac{D}{N}\Psi_1+\sum_{j=2}^N(\frac{D}{N}-w_j)\Psi_j.
\end{align}
For each $l\ne1$, define a dual basis for the $D$ states obtained by omitting $\ket{\Psi_l}$ from the full set $\{\ket{\Psi_j}\}_{j=1}^N$. Denote these bases---one basis for each $l$---as $\ket{\xi_k^{(l)}}$, which satisfy
\begin{align}\label{eqn17}
\inpd{\xi_k^{(l)}}{\Psi_j}=\delta_{jk}~\forall{j,k\ne l}.
\end{align}
Let us first show that $\left\vert\inpd{\xi_k^{(l)}}{\Psi_l}\right\vert=1$. Recalling the comment in the sentence following \eqref{eqn3}, we have
\begin{align}\label{eqn18}
0=\sum_{j=1}^Ne^{2\pi \textrm{i}j/N}\ket{\Psi_j}.
\end{align}
Since $\ket{\xi_k^{(l)}}$ is orthogonal to $\ket{\Psi_j}~\forall{j\ne k,l}$, then multiplying \eqref{eqn18} from the left by $\bra{\xi_k^{(l)}}$ yields
\begin{align}\label{eqn19}
0=e^{2\pi \textrm{i}k/N}+e^{2\pi \textrm{i}l/N}\inpd{\xi_k^{(l)}}{\Psi_l},
\end{align}
and the desired result follows immediately. Recalling that $\Pi_f\ge0$, we now have from \eqref{eqn16} that
\begin{align}\label{eqn20}
0&\le\bra{\xi_k^{(l)}}\Pi_f\ket{\xi_k^{(l)}}=\frac{D}{N}-w_k+(\frac{D}{N}-w_l)\left\vert\inpd{\xi_k^{(l)}}{\Psi_l}\right\vert^2,\notag\\
\end{align}
or
\begin{align}\label{eqn21}
w_k+w_l&\le\frac{2D}{N},~\forall{k,l\ne1},
\end{align}
a result we will use below.

We now turn to the failure probability,
\begin{align}\label{eqn22}
Pr(f)&=\frac{1}{D}\sum_{j=2}^N\Tr{\Pi_f\Phi_j},\notag\\
	&=\frac{1}{D}\sum_{j=2}^N\Tr{\left[I-\sum_{k=2}^Nw_k\Psi_k\right]\Phi_j},\notag\\
	&=1-\frac{1}{D}\sum_{j=2}^Nq_jw_j,
\end{align}
where we have used \eqref{eqn5} followed by \eqref{eqn4}, and defined $q_j=\left\vert\inpd{\Phi_j}{\Psi_j}\right\vert^2$. The $q_j$ can be found by taking the inner product of $\ket{\Phi_k}$ with \eqref{eqn18}, obtaining
\begin{align}\label{eqn23}
0=e^{2\pi \textrm{i}/N}\inpd{\Phi_k}{\Psi_1}+e^{2\pi \textrm{i}k/N}\inpd{\Phi_k}{\Psi_k},
\end{align}
which gives
\begin{align}\label{eqn24}
q_k=\left\vert\inpd{\Phi_k}{\Psi_1}\right\vert^2.
\end{align}
On the other hand, multiplying \eqref{eqn2} by $\Phi_k$ and taking the trace, we have
\begin{align}\label{eqn25}
\frac{N}{D}&=\left\vert\inpd{\Phi_k}{\Psi_1}\right\vert^2+\left\vert\inpd{\Phi_k}{\Psi_k}\right\vert^2,\notag\\
	&=2q_k,
\end{align}
having used \eqref{eqn24} to obtain the last line. Hence,
\begin{align}\label{eqn26}
q_k=\frac{N}{2D}~\forall{k=2,\ldots,N}.
\end{align}
Inserting this into \eqref{eqn22}, we have
\begin{align}\label{eqn27}
Pr(f)=1-\frac{N}{2D^2}\sum_{j=2}^Nw_j.
\end{align}

Let us now see what happens if $\exists k(w_k>D/N)$. Then,
\begin{align}\label{eqn28}
\sum_{j=2}^Nw_j&=\sum_{2=j\ne k}^Nw_j+w_k\notag\\
	&\le \sum_{2=j\ne k}^N(\frac{2D}{N}-w_k)+w_k\notag\\
	&=(N-2)\frac{2D}{N}-(N-3)w_k\notag\\
	&<(D-1)\frac{2D}{N}-(D-2)\frac{D}{N}=\frac{D^2}{N},
\end{align}
where we used \eqref{eqn21} to obtain the second line, and the fact that $D=N-1$ to get the last inequality. Therefore, the maximum of this sum has $w_j=D/N~\forall{j=2,\ldots,N}$, because in this case
\begin{align}\label{eqn29}
\sum_{j=2}^Nw_j&=\frac{D^2}{N}.
\end{align}
Maximizing this sum minimizes $Pr(f)$, see \eqref{eqn27}, and since by \eqref{eqn2} the latter choice of $w_j$ is a valid complete measurement---having measurement operators $\{D\Psi_j/N\}_{j=1}^N$, with $\Pi_f=D\Psi_1/N$---this is therefore our optimal global measurement. From \eqref{eqn27}, we see that this measurement achieves
\begin{align}\label{eqn30}
Pr(f)=0.5,
\end{align}
which is thus our optimal probability of failure. As each $\Psi_j$ is a product operator, this measurement is clearly separable. By Corollary~\ref{cor1} along with the fact, just demonstrated, that there is one and only one set of $\{w_j\}$ that minimizes $Pr(f)$, it is also the unique optimal measurement whose action is restricted to $\HC$. Since every operator in this measurement is a tensor product of positive, rank-$1$ operators on $P$ parties, and noting that rank-$1$ positive operators are extreme rays in the convex cone of all positive operators and therefore must be extreme in the convex cone generated by any set of positive operators, each local part $\psi_j^{(\alpha)}$ of each $\Psi_j$ is extreme in the collection of all $\psi_j^{(\alpha)}$,  for each party $\alpha$. We thus see that $e_\alpha=N~\forall{\alpha}$, and the sum of extreme rays yields $\sum_\alpha e_\alpha=PN>2(N-1)$, an extensive violation of Theorem~\ref{thm1}. Therefore, it is not possible to implement this measurement by finite-round LOCC acting on $\HC$. 

By Lemma~\ref{lem2} we see that for these input states and any LOCC measurement on an enlarged Hilbert space, there is an effectively identical LOCC measurement on $\HC$. The phrase ``effectively identical" means that the two measurements have the same set of outcomes (excluding those outcomes that can never occur, see the appendix), where each outcome in the measurement on the enlarged space has the same probability (and output state) as the corresponding outcome in the other measurement, which is a measurement that acts only on $\HC$. Therefore, if there is a finite-round LOCC measurement on the enlarged Hilbert space that achieves the optimal probability of success, then its ``effectively identical" counterpart, which achieves the same probabilities, is a finite-round LOCC measurement on $\HC$ that also achieves that optimal probability of success. This is a contradiction, since we've just seen that no such measurement on $\HC$ exists, implying there is no such LOCC measurement on the enlarged Hilbert space, either.\footnote{Note that since Theorem~\ref{thm1} only provides a \textit{necessary} condition for LOCC, it may well be the case that the measurement on $\HC^\prime$ satisfies the bound in that theorem, even when that measurement cannot be implemented by LOCC.} We therefore see that no finite-round LOCC protocol can be optimal, including those that act on an enlarged Hilbert space, and this completes the proof of Theorem~\ref{thm2}.


\section{Conclusions}\label{sec4}
We have presented a class of problems involving the unambiguous discrimination of quantum states, and have shown in Theorem~\ref{thm2} that for each element in this class, there exists an optimal, separable measurement, achieving the minimum possible failure probability of $0.5$, which is the unique such measurement that acts only on the space $\HC$ spanned by the set of states to be discriminated. We then demonstrated the utility of Theorem~\ref{thm1} of \cite{myExtViolate1}, a recently discovered necessary condition that a separable measurement can be implemented by finite-round LOCC, by using the latter theorem to (easily) prove that this separable measurement cannot be implemented by LOCC in any finite number of rounds. Finally, we showed that any LOCC measurement on an enlarged Hilbert space must also be strictly less than optimal. We note that this class of problems is infinitely large, having at least one element for each prime number $N$. (Generally it will, in fact, have more than one element for any given $N$, as long as $D=N-1$ is not the product of two primes.) Therefore, we have solved an infinite set of unambiguous discrimination problems, each of which has an optimal measurement that is separable, but for which there is no finite-round LOCC measurement that is optimal. Due to a result of \cite{Erdos2}, this class of problems includes an infinite number of examples for each number of parties, $P$, and we have included an explicit example here for the simplest system of two qubits, see \eqref{eqn12}.

If the parties are given multiple copies of the chosen state, 
\begin{align}\label{eqn31}
\ket{\Phi_j^{\otimes n}}=\stackrel{n~\textrm{copies}}{\overbrace{\ket{\Phi_j}\otimes\cdots\otimes\ket{\Phi_j}}},
\end{align}
then one can use the same arguments used above for a single copy to also show that there exists an optimal global measurement that is separable and that achieves the minimum possible $Pr(f)=2^{-n}$. This optimal measurement consists of the $N^n$ measurement operators $\{(D/N)^n\Psi_{j_1}\otimes\Psi_{j_2}\otimes\cdots\otimes\Psi_{j_n},~j_k=1,\ldots,N~\forall{k}\}$, and one can easily show that this measurement cannot be implemented by finite-round LOCC, again by using Theorem~\ref{thm1}. However, there are now an infinite number of other measurements that act only on the original Hilbert space and also achieve this same $Pr(f)$, and we do not at present know whether any of these are separable, let alone if they are LOCC. It is thus an open question whether or not LOCC is as good as separable measurements for these states in the multiple-copy scenario.

For the single-copy case considered in this paper, we conjecture that the set of states $\SC_\Phi$ cannot be optimally unambiguously discriminated by LOCC even with an infinite number of rounds of communication. We have discussed why we believe this is so in the conclusions of \cite{myExtViolate1}. As an early step toward proving this conjecture, we have recently managed to prove a result which implies the following conclusion about the optimal measurement for this task, $\MC_{opt}$ of Theorem~\ref{thm2}: if there exists an LOCC protocol implementing $\MC_{opt}$, then every branch of this protocol must continue for an infinite number of rounds. That is, in any such protocol, no outcome of any intermediate measurement can be terminal---the parties must continue measuring forever no matter what outcomes have been obtained in earlier rounds. While this result does not in itself prove the conjecture, it does strengthen our belief that $\MC_{opt}$ cannot be implemented by LOCC even with an infinite number of rounds, and we hope to find a full proof of this result in the not-too-distant future.

\noindent\textit{Acknowledgments} --- The author would like to thank Li Yu for helpful comments, and an anonymous referee for asking about the possibility of measurements on an enlarged Hilbert space, a question which led directly to Lemma~\ref{lem2}. This work has been supported in part by the National Science Foundation through Grant No. 1205931.
\appendix*
\section{Proof of Lemma~\ref{lem2}}
Consider any measurement on the enlarged Hilbert space $\HC^\prime$ and 
suppose the Kraus operators for that measurement are the set $\{K_j\}$. Define $\Pi=\Pi_1\otimes\Pi_2\otimes\ldots$ to be the projector (from $\HC^\prime$) onto the original system $\HC$. Then, the set of Kraus operators, $\{K_j\Pi,I^\prime-\Pi\}$, is a complete measurement on $\HC^\prime$, where $I^\prime$ is the identity operator on $\HC^\prime$. 
There is no guarantee that $I^\prime-\Pi$ is a product operator, but one can write
\begin{align}\label{eqnA1}
I^\prime-\Pi= (I^\prime_1-\Pi_1)\otimes I^\prime_2\otimes\ldots\otimes I^\prime_P &+ \Pi_1\otimes(I^\prime_2-\Pi_2)\otimes I_3^\prime\otimes\ldots\otimes I^\prime_P+ \ldots \notag\\
									& + \Pi_1\otimes\Pi_2\otimes\ldots\otimes\Pi_{P-1}\otimes(I^\prime_P-\Pi_P),
\end{align}
which as we will see in a moment is a sum of product operators that, along with $\Pi$, can be implemented by LOCC. Now suppose the measurement under consideration, which acts on the larger space, can be implemented by LOCC using, say, protocol $\LC$. Then, consider the LOCC protocol consisting of protocol $\LC$ preceded by a series of measurements as follows: Party $1$ starts out by doing a two-outcome measurement $\{\Pi_1,I^\prime_1-\Pi_1\}$. If she gets the second outcome, they terminate the protocol, but otherwise party $2$ measures $\{\Pi_2,I^\prime_2-\Pi_2\}$. If he gets the second outcome, they terminate, but otherwise party $3$ measures $\{\Pi_3,I^\prime_3-\Pi_3\}$, and so on until all parties have done this, after which they proceed with protocol $\LC$. Now, since we consider only input states supported on $\HC$, $I^\prime_1-\Pi_1$ has zero probability of occurrence, as do the other outcomes $I^\prime_\alpha-\Pi_\alpha$ for each party $\alpha$. Under these circumstances, the modification of protocol $\LC$ becomes an LOCC measurement acting on the original space $\HC$ alone, which has the exact same probabilities as does protocol $\LC$, and in fact, the output state for each branch of the protocol is also identical in the two cases. 

Or perhaps it is more precise to put it this way: Since outcomes $I^\prime_\alpha-\Pi_\alpha$ have zero probability of success, nothing changes if we simply begin at the point where $\Pi$ has been implemented (after all $P$ parties have performed their two-outcome measurements  $\{\Pi_\alpha,I^\prime_\alpha-\Pi_\alpha\}$). From a purely technical perspective, we can't really do this on the enlarged space, because then the POVM elements don't represent a complete measurement, but we can certainly do this on the original space instead of the enlarged one. Then this is an LOCC measurement on $\HC$, which achieves the exact same result as does protocol $\LC$. Indeed in a matrix representation, we can write $\Pi=\textrm{diag}\left(I_\HC,0_{\HC^\perp}\right)$ and 
\begin{align}\label{eqnA2}
K_j\Pi=\left(\tilde K_j~~\tilde0\right),
\end{align}
where $I_\HC$ is the identity operator on $\HC$, $\tilde K_j$ is an operator that acts only on $\HC$, and $0_{\HC^\perp}$ and $\tilde0$ are zero operators that act only on $\HC^\perp$, which we define to be the orthogonal complement of $\HC$ in $\HC^\prime$. Note that whereas $0_{\HC^\perp}$ is a square matrix, $\tilde K_j$ and $\tilde0$ need not be square; they may map to a space larger or smaller than that on which they act, but both map to the same output space, say $\HC_{out}$, which is also the output space for the operator on $\HC^\prime$ that we started with, $K_j$. Then the matrix $\tilde K_j$ represents an operator that acts on $\HC$ (and whose output is $\HC_{out}$), and the collection of these operators constitutes a measurement $\{\tilde K_j\}$ acting only on $\HC$, which is effectively identical to the measurement $\{K_j\}$ when the latter acts only on inputs that are confined to $\HC$. That is, measurement $\{\tilde K_j\}$, which is a measurement that acts only on $\HC$, achieves exactly what is achieved when $\{K_j\}$ acts on inputs that are confined to $\HC$. This completes the proof.\endproof


\end{document}